\documentclass[prb,twocolumn,draft,amsmath,showpacs]{revtex4}
\usepackage{graphics}

\begin{document}

\bibliographystyle{prsty}
\input epsf
\def\ns2{NbSe$_2$}

\title {Unusual magnetic field-induced phase transition in the mixed state of
superconducting NbSe$_{2}$}

\author {A. V. Sologubenko$^1$, I. L. Landau$^{1,2}$, H. R. Ott$^1$, A.
Bilusic$^{1,3}$, A. Smontara$^3$, and H. Berger$^4$}

\affiliation{$^1$Laboratorium f\"ur Festk\"orperphysik, ETH H\"onggerberg,
CH-8093 Z\"urich, Switzerland}

\affiliation{$^2$Kapitza Institute for Physical Problems, 117334 Moscow,
Russia}

\affiliation{$^3$Institute of Physics, POB304, HR-10001 Zagreb,
Croatia}

\affiliation{$^4$Institut de physique de la mati\'ere complexe, EPFL, CH-1015 Lausanne, Switzerland}

\date{\today}

\begin{abstract}
    
The thermal conductivity $\kappa$ in the basal plane of single-crystalline 
hexagonal NbSe$_2$ has been measured as a function of magnetic field $H$, 
oriented both along and perpendicular to the $c$-axis, at several 
temperatures below $T_{c}$. With the magnetic field in the 
basal plane and oriented parallel to the heat flux we observed, in 
fields well below $H_{c2}$, an unexpected hysteretic behavior of 
$\kappa(H)$ with all the generic features of a first order phase
transition. The transition is not manifest in the $\kappa(H)$ 
curves, if $H$ is still in the basal plane but oriented perpendicularly 
to the heat-flux direction. The origin of the transition is not yet 
understood.

\end{abstract}
\maketitle

The layered hexagonal compound \ns2 has been of interest to researchers 
for some time, mainly because of the relatively strong  anisotropy of its 
superconducting properties. Pure samples exhibit a superconducting critical 
temperature $T_c \approx 7.2$ K, and the upper critical fields for $H$ along 
and perpendicular to the hexagonal $c$-axis are $H_{c2}^{c} = 40$~kOe and  
$H_{c2}^{ab} = 140$~kOe,\cite{Woollam76} respectively. The anisotropy 
of the electron system is also reflected in the charge density wave
(CDW) transition at $T_{\rm CDW} = 34 {\rm ~K}$ (in pure \ns2),
interesting by itself and also in view of the competition between 
the CDW formation and superconductivity. \cite{Sooryakumar80} 
Another important feature of  \ns2  is the existence of a variety of 
vortex matter related phenomena, such as a pronounced 
peak effect and various phase transitions and instabilities (see, e.g., 
Refs.~\onlinecite{Banerjee99_Dis,Marchevsky01,Paltiel00_Ins} and references therein).

In this paper, we present the observation of an unexpected hysteretic
behavior of the thermal conductivity  in external magnetic 
fields $\kappa$ with the characteristic features of
a first order phase transition.   Several anomalies
associated with the transition are yet to be explained.

The single crystal used in the experiments was grown by employing the iodine
vapor transport method as described in Ref.~\onlinecite{Levy83}. Two bar-shaped
samples with dimensions of $4.4 \times 0.88 \times 0.42 {\rm ~mm}^{3}$ and
$2.8 \times 0.51 \times0.50 {\rm ~mm}^{3}$, were cut  from the crystal.
Very detailed measurements of $\kappa(T,H)$ were done on one of the
samples and the other specimen was used to ascertain that the 
observed effects described below are indeed sample-independent. The thermal 
conductivity was measured by employing the standard uniaxial heat flux 
method in an experimental setup that was also used in our recent study 
of $\kappa(T,H)$ in MgB$_{2}$.\cite{Sologubenko02_Tem,Sologubenko02_The}
The heat flux $\dot{\bf Q}$ was kept in the basal plane and along the longest 
lateral dimensions of the sample. The temperature difference $\Delta 
T$ along the specimen was kept to about 3 - 5 \% of the average sample 
temperature. The zero-field superconducting transition temperature 
$T_{c}=7.2$~K was  determined via four-contact $dc$ resistivity measurements 
and from the results of measurements of the magnetic susceptibility, 
employing a commercial Quantum Design SQUID magnetometer.

\begin{figure}[t]
   \begin{center}
    \leavevmode
    \epsfxsize=1\columnwidth \epsfbox {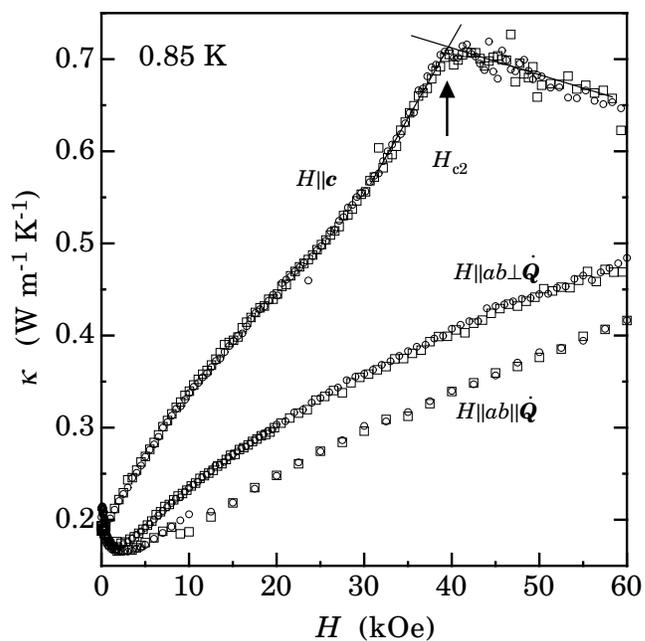}
     \caption{
     The basal plane thermal conductivity of single crystalline \ns2 as 
     a function of differently oriented magnetic fields at 
$T=0.85$~K. The open
     circles and squares correspond to increasing and decreasing fields,
     respectively.
    }
\label{KvsH}
\end{center}
\end{figure}
Fig.~\ref{KvsH} shows $\kappa(H)$ at $T=0.85$~K for 
all three orientations of the magnetic field that we
have explored. Similar results were obtained at other temperatures in 
the interval between 0.38~K and $T_{c}$. The data for $H \parallel c$ 
are in good agreement with the recent results presented by Boaknin {\em et al.}.\cite{Boaknin03}
The main features of $\kappa(T,H)$ can  readily be
understood by considering that the total thermal conductivity
consists of two components, namely the phonon contribution
$\kappa_{\rm ph}$ and the quasiparticle (electron) contribution
$\kappa_{e}$. The normal electronic excitations in the cores of the vortices in the 
mixed state considerably reduce the phonon mean free path, but they 
also add positively to the heat transport.
Hence when the external magnetic field is 
enhanced to above the lower critical field  $H_{c1}$, $\kappa_{\rm 
ph}(H)$ is efficiently reduced but,  simultaneously, $\kappa_{e}(H)$ 
starts to 
increase. The result is a minimum in the $\kappa (H)$ curve in a 
magnetic field slightly exceeding $H_{c1}$. These features may be seen in
Figs.~\ref{KvsH} and 2(b). In the normal state, above the upper critical 
field $H_{c2}$, the thermal conductivity varies only weakly with 
$H$.
It reflects the field induced positive variation of the electrical 
resistivity of NbSe$_2$ in the normal 
state, which also affects the electronic contribution $\kappa_{e}$ to $\kappa$.

\begin{figure}[t]
   \begin{center}
    \leavevmode
    \epsfxsize=1\columnwidth \epsfbox {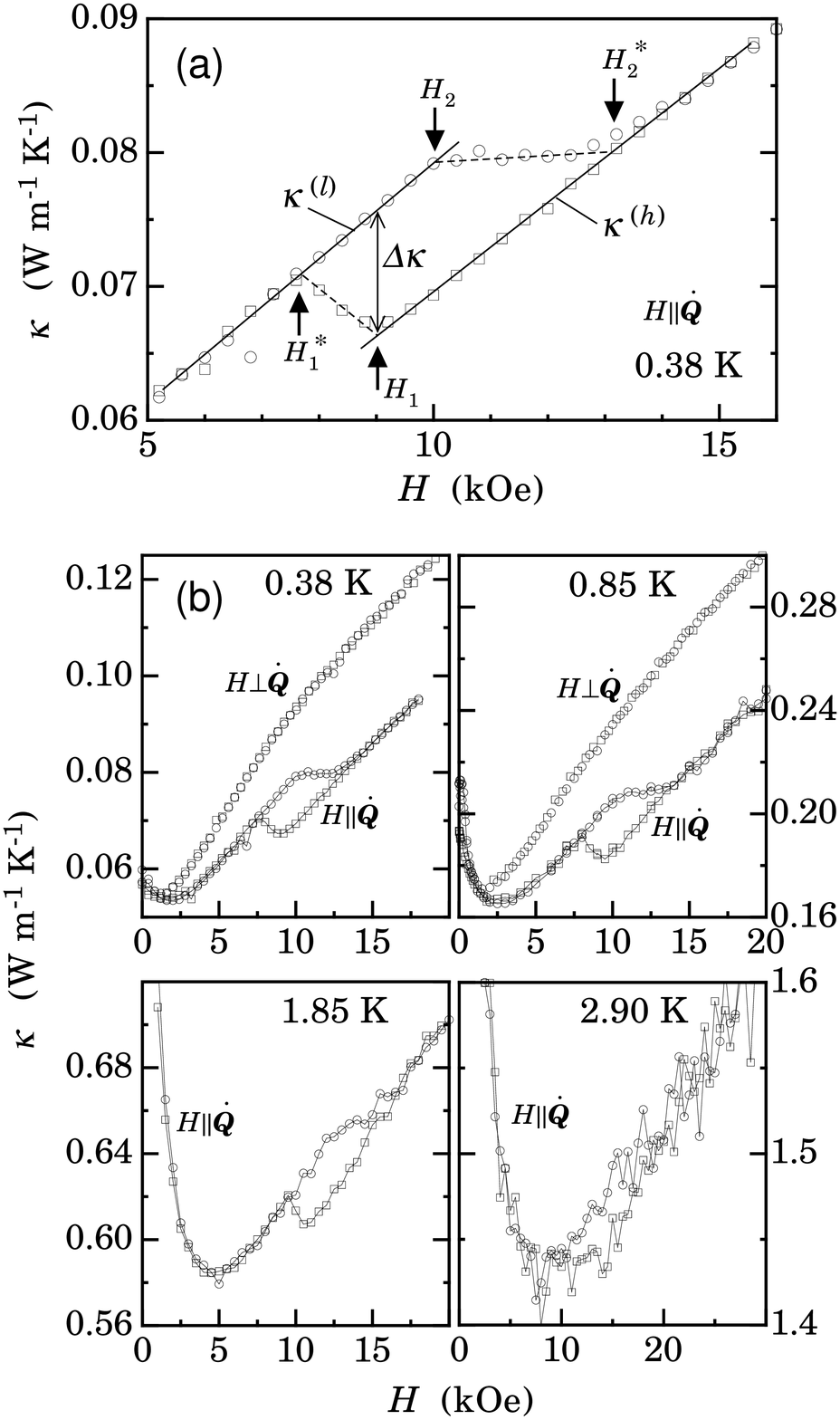}
     \caption{
     (a) Characteristic features of the hysteresis exemplified by
     $\kappa(H)$ at $T = 0.38$~K. The open
     circles and squares correspond to increasing and decreasing field,
     respectively.
     (b) The thermal conductivity as a function of magnetic field oriented
     perpendicularly to the $c$-axis at several temperatures. Note the 
     difference for different field orientations.
    }
\label{Hyst}
\end{center}
\end{figure}

This general
behavior of $\kappa(H)$ is observed for all orientations of the magnetic field, but,
with $H \perp c$ and parallel to the heat-flux direction, 
a pronounced hysteretic behavior in $\kappa(H)$, as shown in 
Fig~\ref{Hyst}, is observed.
This feature, emphasized in Fig.~\ref{Hyst}~(a), is a clear manifestation of some kind of phase
transition, most likely of first order.   
As we
discuss in detail below, some details are not compatible with a conventional 
first order transition, however.
Inspecting the data presented in Fig~\ref{Hyst}, it is clear that
two different $\kappa(H)$ regimes have to be distinguished.
They correspond to two different modifications or phases of the mixed
state, one of which is stable in low fields ($l$-phase) and the other
($h$-phase) corresponds to the equilibrium state in high magnetic fields. 
In the
following we shall use $\kappa^{(l)}(H)$ and $\kappa^{(h)}(H)$ to
denote the thermal conductivities of the $l$- and $h$-phases of the mixed
state, respectively (see Fig. 2(a) for the definitions of $\kappa^{(l)}$, 
$\kappa^{(h)}$ and the corresponding characteristic fields). While
the $l$-phase is stable up to $H = H_2$, the high-field modification 
remains stable down to $H = H_1 < H_2$. Also note that the transition from one 
modification to the other is rather narrow upon field reduction, 
whereas the 
transition in increasing field is more than twice as wide, with 
$\kappa(H)$ being almost constant in the transition region. 
All the features shown in Fig.~\ref{Hyst} were confirmed to be 
reproducible and they  were observed for both investigated samples.

The hysteresis is completely absent if the magnetic field,
while still parallel to the basal plane, is rotated to be perpendicular to the heat flux
direction (see Fig~\ref{Hyst}(b)). This disappearance of the 
hysteresis is quite 
unexpected because, considering the hexagonal symmetry of the  crystal structure of \ns2, 
no anisotropy is expected in the basal plane. It may also be seen  that 
$\kappa(H)$ is notably higher if the magnetic field is oriented 
perpendicularly to the heat flux.

\begin{figure}[t]
   \begin{center}
    \leavevmode
    \epsfxsize=1\columnwidth \epsfbox {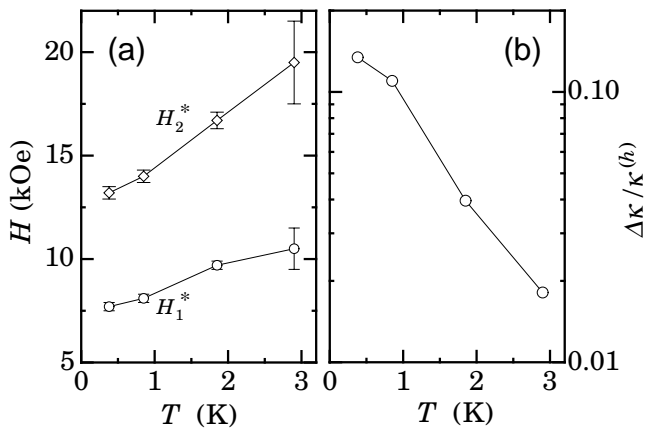}
     \caption{
     (a) Temperature dependencies for the fields $H_1^{*}$ and
     $H_2^{*}$.
     (b) Temperature dependence of the normalized difference $\Delta 
     \kappa = \kappa^{(l)} - \kappa^{(h)}$. For definitions, see 
     Fig.~\ref{Hyst}~(a).
    }
\label{CrFields}
\end{center}
\end{figure}
Another unusual feature of this transition is the considerable width 
of the hysteresis loop with $H^{*}_{2}/H^{*}_{1} \approx 1.7$. It 
turns out that both
$H^{*}_{1}$ and $H^{*}_{2}$ increase with increasing temperature
(see Fig.\ref{CrFields}~(a)).  Considering that all
critical fields of a superconductor usually decrease with increasing
temperature, this is again an unexpected observation.

The magnetic field ranges $H_{2} < H < H_{2}^{*}$ in increasing and $H_{1}^{*} < H <
H_{1}$ in decreasing fields, respectively, correspond
to transition regions where the $l$- and the $h$-phase coexist. 
Because both $\kappa^{(l)}(H)$ and $\kappa^{(h)}(H)$ may quite well be
approximated by parallel straight lines, the volume ratio between the two 
phases for each point in the transition regions can easily be evaluated 
from the experimental data.

\begin{figure}[t]
   \begin{center}
    \leavevmode
    \epsfxsize=1\columnwidth \epsfbox {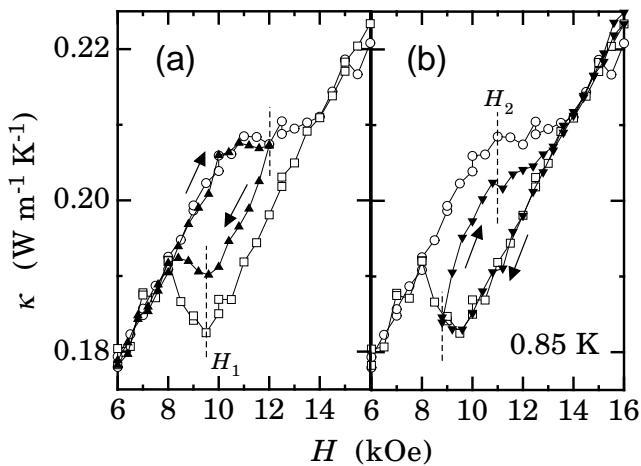}
     \caption{
     Partial hysteresis loops of $\kappa(H)$ at $T=0.85$~K. The open
     circles and squares represent the full loop and are the same as 
     in Fig.~\ref{Hyst}(b).
      (a) $H$ is first increasing from 0 to 12~kOe and then decreasing.
      (b) $H$ is first enhanced from 30~kOe to 8.8~kOe and then 
      reduced. The broken vertical lines mark the special fields 
      values.
    }
\label{Loops}
\end{center}
\end{figure}
\begin{figure}[t]
   \begin{center}
    \leavevmode
    \epsfxsize=0.9\columnwidth \epsfbox {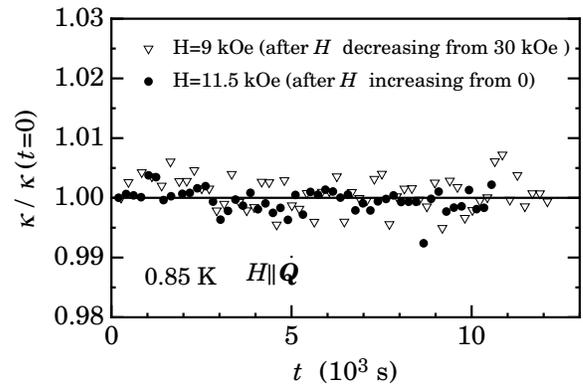}
     \caption{
     $\kappa$ vs time for two values of the external magnetic field 
     at $T = 0.85 K$.
    }
\label{TimeDep}
\end{center}
\end{figure}
Partial hysteresis loops, as shown in Fig.~\ref{Loops}, are also
unusual. If a reduction of the magnetic field is started from the transition
region between $H_{2}$ and $H^{*}_{2}$ (Fig.~\ref{Loops}~(a)), the 
measured $\kappa (H)$ curve is practically parallel to $\kappa^{(l)}(H)$ 
and $\kappa^{(h)}(H)$ until $H_{1}$ is reached. Roughly the same 
feature emerges if the magnetic field is enhanced from the transition
region $H_{1} > H > H^*_{1}$ (see Fig. 4(b)). These observations imply 
that in the limited field range between the vertical broken lines in 
Fig. 4, the volume ratio between the $l$- and the $h$-phase in the sample 
is independent of the applied magnetic field and can only be changed if 
the magnetic field exceeds $H_{2}$
(in increasing field) or is reduced to below $H_{1}$ (in decreasing
field). This behavior is rather difficult to understand in terms of a 
common first order phase transition. Indeed, in conventional first order 
transitions, the main reason for a hysteresis is the inertia toward 
generating a sufficiently large nucleus of a new phase. Once the nucleus 
exists, the transition usually happens in a rather narrow range 
of fields or temperatures. Here we have a completely different situation. 
Two different phases coexist in fixed amounts over an extended field 
region and it needs to be understood why
the transformation from one phase to the other does not happen upon varying 
the applied magnetic field within the limits mentioned above.

We also tested the stability of $\kappa$ at fixed values of the magnetic 
field in the transition regions and at constant temperature. As may be 
seen in Fig.~\ref{TimeDep}, there are no observable changes of $\kappa$ 
with time. This observation provides again clear evidence that the 
transformation of one phase into the other cannot happen as easily as it 
usually does in the case of common first order phase transitions. In 
an extended range of the magnetic field both the $l$- and the $h$-phase 
are present and stable in the sample. Taking into account that near the edges of the 
hysteresis loop only one of them represents the equilibrium, the stability 
of the described situation implies that some potential
barrier seems to prevent the completion of the transition from one 
equilibrium phase to the other.

Before we discuss the situation further, we note that at temperatures as low
as 0.38~K, the phonon contribution $\kappa_{\rm ph}$ to the thermal conductivity
in the mixed state is small compared to $\kappa_e$, certainly in 
fields beyond the $\kappa(H)$ minima (see Figs.~\ref{KvsH} and 
\ref{Hyst}). Because we are mainly
interested in low magnetic fields $H\le 0.1H_{c2}$, no overlap of vortex
cores needs to be considered and each vortex line is expected to contribute
to the heat transport individually. If the magnetic field is perpendicular 
to the heat flux, the vortex motion along the heat flux, induced by a 
temperature gradient $\nabla T$, also contributes to $\kappa$, but this 
contribution is not expected to be significant (see Ref.~\onlinecite{Freimuth03} 
and references therein).

The results presented above clearly imply the existence of a phase transition 
in the mixed state of our \ns2 samples with the following main features:

(i)~~A very large difference $\Delta \kappa = \kappa^{(l)} - \kappa^{(h)}$ 
between the thermal conductivities in the low and the high field phase.
If we assume that $\Delta \kappa$ is only due to a change of the
vortex density, we have to assume an unrealistically large variation of
the magnetic induction $\Delta B \approx 3$ kG (see Fig.~\ref{Hyst}~(a)).

(ii)~~A very large width of the hysteresis loop with $H_2^*/H_1^* = 1.7$.

(iii)~~The coexistence of the $l$- and the $h$-phase of the mixed state in 
an extended range of magnetic fields. The adopted volume ratio is 
practically independent of $H$ in this range (Fig.~\ref{CrFields}). 
This particular feature is incompatible with a conventional first order 
transition.

(iv)~~A completely different behavior of $\kappa(H)$, if the magnetic
field is oriented perpendicularly to the heat flux, exhibiting no visible 
signs of a phase transition (Fig.~\ref{Hyst}~(b)).

(v)~~A shift of the transition to higher fields with increasing
temperature (Figs.~\ref{Hyst}~(b) and \ref{CrFields}~(a)).

Although the phase transition manifests itself by a rather prominent
feature in the $\kappa (H)$ curves, we could not find a reasonable scenario
which would explain all the different features of the transition. Below 
we briefly consider several possible phase transitions in the mixed state of 
type-II superconductors and discuss their  relevance for our 
experimental observations.

1.~~The well known vortex-lattice melting transition explains
neither the magnitude of the observed difference $\Delta \kappa =
\kappa^{(l)} - \kappa^{(h)}$, nor the vanishing of the transition for $H \perp
\dot{\bf Q}$. Quite unlikely are other vortex ordering transitions
considered in the literature, because all of them invoke only
small changes of the vortex density. We also recall that the vortex
lattice melting is a typical first order transition and therefore it
is incompatible with the partial hysteresis loops shown in Fig.~\ref{Loops}.

2.~~It was pointed out that in superconductors with other ordering 
phenomena, a first order phase transition may occur due to the competition
between superconductivity and the additional ordering phenomenon. \cite{Kurin94} 
As an example of additional ordering, the authors of Ref.~\onlinecite{Kurin94} 
considered a charge density wave formation. This matches exactly the case for NbSe$_2$ but, because 
the assumed transition is of  first order,  it is not clear 
how the coexistence 
of two mixed state phases and the dependence of the $\kappa(H)$ feature on the 
orientation of the magnetic field can be explained.

3.~~The first order transition discussed in Ref.~\onlinecite{Paltiel00_Ins} 
was observed for $H \parallel c$ and therefore seems not to be 
relevant to our case. The same applies to other reports of 
instabilities of the mixed state of \ns2 which were observed either 
close to $H_{c2}$ or for other field orientations.\cite{Banerjee99_Dis,Marchevsky01}

4.~~The possible phase transition discussed in 
Ref.~\onlinecite{Landau02} is expected 
to occur in fields rather close to $H_{c2}$ and therefore is {\it a priori} 
considered as unlikely to be the reason for our observation.

In conclusion, measurements of the thermal conductivity in external 
magnetic fields revealed an unusual phase transition in the mixed state
of NbSe$_2$. The origin and the nature of this transition is not yet understood 
and further experiments are needed to clarify the situation.

\acknowledgments

This work was financially supported in part by
the Schweizerische Nationalfonds zur F\"orderung der Wissenschaftlichen
Forschung. A.B. acknowledges financial
support from the Swiss Federal Commission for Foreign Students.

\end{document}